# First principles study of the graphene/Ru(0001) interface

De-en Jiang[1,*], Mao-Hua Du[2], and Sheng Dai[1]

[1]Chemical Sciences Division and [2]Materials Science and Technology Division, Oak Ridge National Laboratory, Oak Ridge, Tennessee 37831

Annealing the Ru metal that typically contains residual carbon impurities offers a facile way to grow graphene on Ru(0001) at the macroscopic scale. Two superstructures of the graphene/Ru(0001) interface with periodicities of 3.0-nm and 2.7-nm, respectively, have been previously observed by scanning tunneling microscopy. Using first-principles density functional theory, we optimized the observed superstructures and found interfacial C-Ru bonding of C atoms atop Ru atoms for both superstructures, which causes the graphene sheet to buckle and form periodic humps of ~1.7 Å in height within the graphene sheet. The flat region of the graphene sheet, which is 2.2-2.3 Å above the top Ru layer and has more C atoms occupying the atop sites, interacts more strongly with the substrate than does the hump region. We found that interfacial adhesion is much stronger for the 3.0-nm superstructure than for the 2.7-nm superstructure, suggesting that the former is the thermodynamically more stable phase. We explained the 3.0-nm superstructure's stability in terms of the interplay between C-Ru bonding and lattice matching.

*To whom correspondence should be addressed. E-mail: jiangd@ornl.gov. Phone: (865)574-5199. Fax: (865) 576-5235.

I. INTRODUCTION

Successful isolation of the graphite monolayer, graphene, in 2004 has opened up a new field of research.[1,2] Interesting new physics has been discovered for graphene's two-dimensional lattice[3-7] and novel devices based on graphene have been made.[8] Applications of graphene for electronic devices in a large scale demand a way to produce large quantity of graphene cheaply and reliably. The original method of mechanical exfoliation[1] is not quite suited for large-scale production. Hence, alternative methods to produce graphene have been developed. One attractive method is to grow graphene on a substrate, such as SiC[9], Ir[10,11], and Ru.[12-16]

The method to grow graphene on Ru(0001) is especially promising in that one can just anneal a commercially available Ru sample that typically has residual carbon impurities. During annealing, the C atoms will segregate onto the Ru surface and form a macroscopic graphene layer on Ru(0001), up to millimeter-scale.[16] It has been shown that this graphene layer is stable in ambient conditions and can survive high temperature treatment.[12] It has also been found that a second layer of graphene can grow over the first layer after the first layer's completion and this second layer is weakly coupled to the substrate.[15] At the microscopic level, a hexagonal superstructure with a repeating length of ~3.0 nm has been observed by scanning tunneling microscopy (STM), and attributed to be a moiré pattern caused by matching (12×12) graphene primitive cells to (11×11) Ru(0001) primitive cells.[12,13,15,16] A smaller superstructure with a repeating length of 2.7 nm has also been observed by another STM study.[14]

At the atomic level, different results regarding the interfacial structures and interactions have been reported for the 3.0-nm superstructure of graphene on Ru(0001). Marchini et al. measured an apparent height of 1.8 Å for graphene on Ru(0001)[12] by STM, but Sutter et al. determined an optimal height of 1.45 Å by fitting the experimental I(V) curve from low-energy electron microscopy (LEEM).[15] The experimental evidence from STM, LEEM, and the work function measurements suggests a strong interaction at the graphene/Ru interface.[12,15] However, a preliminary density functional theory (DFT) study of the 3.0-nm superstructure showed that the interfacial interaction is weak and the graphene layer has a corrugation of only 0.4 Å and is at least 3.9 Å above the Ru substrate.[16] Another DFT study by Wang et al. found alternating strong and weak interaction at the graphene/Ru(111) interface and a corrugation of 1.5 Å within the graphene sheet.[17] Wang et al. also computed the STM image for the 3.0-nm superstructure which shows good agreement with experiment.

In contrast to the 3.0-nm superstructure which has been reported in many recent studies, the smaller 2.7-nm superstructure has attracted less attention. Although patterns of graphene on Ru(0001) depend on the growth conditions,[16] the fact that there are more reported cases of the 3.0-nm superstructure seems to indicate that it may be the thermodynamically stable phase. However, from a pure geometrical point of view, the 2.7-nm superstructure corresponds to a better match between the graphene and Ru(0001) lattices. This apparent conflict calls for an explanation which may be obtained by first-principles calculations. In this paper, we will show that the 3.0-nm superstructure is indeed preferred over the 2.7-nm one. We will also explain why that is the case, by examining structures, bonding, and adhesion energetics for the two superstructures. Our results support the conclusion of the earlier DFT study in Ref. 17 that there is strong binding between the graphene sheet and the Ru surface.

**II. COMPUTATIONAL METHODS**

We employed the periodic density functional theory with a plane wave basis (kinetic energy cutoff, 400 eV).[18,19] For electron exchange-correlation, the Perdew-Burke-Ernzerhof (PBE) form[20] of generalized-gradient approximation (GGA) was used. Although DFT-PBE cannot describe well the dispersion interaction between graphene layers, it yields quite reasonable results for the interaction between graphene and Ru(0001), as we show later. Moreover, previous studies[21,22] showed that DFT-GGA should be able to capture well the interaction between the benzene π-system and transition metals in the iron group. To describe the electron-core interaction, we utilized the projector-augmented wave method within the frozen-core approximation.[23,24] Optimized lattice parameters for graphene (a=2.465 Å) and Ru (hcp structure: a=2.724 Å, c=4.308 Å) agree very well experiment. The superstructures of the graphene/Ru(0001) interface were modeled by using the super cell approach. Our models of the superstructures include the graphene layer, three layers of Ru, and a 12-Å vacuum layer. Only the Γ point was used to sample the Brillouin zone, and the force tolerance for structural optimization was set at 0.05 eV/Å.

### III. RESULTS AND DISCUSSION

We started by examining the 3.0-nm superstructure which has been reported by several independent studies.[12,13,15,16] This superstructure contains a (12×12) lattice of repeating graphene primitive cells matched to a (11×11) lattice of repeating Ru(0001) primitive cells, with the [10$\bar{1}$0] directions of the two lattices aligned. We initially placed the flat graphene sheet close to the top Ru layer at a height of 1.9 Å to test the hypothesis of chemical interaction between graphene and Ru(0001). This initial distance is much shorter than one half of the sum of graphite and Ru(0001)'s interlayer spacings (2.8 Å) and slightly higher than the apparent height (1.8 Å) observed by STM.[12] After structural relaxation, we found that a hump rises from the otherwise relatively flat graphene sheet [Fig. 1(a)]. The height of the hump is about 1.67 Å, larger than the average corrugation (0.7 – 1.1 Å) observed by STM which we note is subject to bias dependence.[12] The optimized structure shows a height of ~2.2 Å over the top Ru layer for the low, flat region of the graphene sheet. This distance matches well the interlayer distance between Ru layers of Ru(0001) (at 2.15 Å), which may help explain the observation that graphene grows at the lower terrace of a step on Ru(0001),[12,15] as a way to extend the edge of the upper terrace. Fig. 2 is a perspective view of the humps in the graphene layer, which offers a clear contrast between the corrugations and the lateral dimensions. We found relaxation in the Ru substrate small and the closed-packed geometry of Ru(0001) well preserved. The buckling in the top Ru layer is found to be ~0.2 Å and the lateral relaxation of Ru atoms is even smaller (< 0.1 Å).

Our DFT results for the 3.0-nm superstructure agree very well with Ref. 17 on the strong graphene-Ru binding and a significant buckling of the graphene layer, but disagree with Ref. 16, which showed a weak binding with minimum interfacial spacing of 3.9 Å and a small corrugation (< 0.4 Å) within the graphene layer.

Fig. 1(b) shows the top view of the optimized structure with the height of the C atoms color-coded. One can see that in the low, flat region (at the lower left and upper right areas) where graphene is close to the Ru layer and supposed to interact more strongly (than the hump region) with the substrate, more C atoms sit atop Ru atoms. This is opposite to the structural model used in the LEEM experiment,[15] which assumed that all the carbon atoms occupy the hollow sites on the Ru(0001) surface. In contrast to the flat region, the carbon atoms in the hump region [see the center region of Fig. 1(b)] are indeed over the hollow sites of the Ru (0001) surface.

Now we discuss the energetics at the graphene/Ru(0001) interface. We found an adhesion energy of 6 eV per superstructure (21 meV/C-atom or 0.12 J/m$^2$) between graphene and Ru(0001), in good agreement with the 6.7-eV adhesion energy found by Wang et al.[17] Given that about a third of the C atoms are in the hump region and interact weakly with the Ru substrate, this strength is comparable to an interfacial strength of 30 meV/C-atom (obtained by using DFT-PBE as well) found for the graphene/Ni(111) interface where the unbuckled graphene layer (due to the nearly perfect lattice match with the substrate) is 2.13 Å above the Ni substrate.[25]

In Fig. 3 we plot the electron density difference between the interface and a sum of the separate graphene and Ru(0001) surfaces. One can see that charge depletes from the top Ru layer and accumulates in the interfacial region near the graphene layer. A zoom into the flat region of the graphene layer in Fig. 3(a) shows that the accumulated charge localizes along the C-Ru bonds. The top view [Fig. 3(b)] shows that there is little charge transfer between the hump region of the graphene sheet and Ru(0001) due to their large separation. A closer look into the flat region of the graphene layer [upper right corner of Fig. 3(b)] shows that (1) charge transfer is concentrated along sites where C atoms sit atop Ru atoms and (2) there is little charge transfer on the sites where C atoms are located over the hollow sites even though these C atoms are also in the flat region and close to the top Ru layer. This again supports the conclusion that the C-Ru bonding arising from the C atoms at the atop sites dominates the graphene/Ru(0001) interfacial adhesion. To further illustrate this point, Fig. 4 displays the orbital-decomposed local density of states for a Ru atom and the C atom on top of it [as indicated by a straight arrow in the zoomed-in region of Fig. 3(b)]. One can clearly see strong hybridization between C $2p_z$ and Ru $4d_z^2$ orbitals which are normal to the interface.

The measured work functions for the bare and the graphene-covered Ru(0001) surfaces are 5.4 and 4.5 eV, respectively, comparable to the calculated values of 5.3 and 3.5 eV. The large work function change upon the graphene adsorption suggests significant charge transfer at the graphene/Ru interface, in consistent with our calculations. However, the adsorbate-induced decrease of the work function usually results from the charge transfer from the adsorbate to the substrate, opposite to that shown in Fig. 3. Such "abnormal" work function change has been observed by several authors in their studies of surface adsorption,[26-28] and has been argued to arise from the charge transfer within the adsorbates that counteracts the charge transfer from the substrate to the adsorbates. Note that Fig. 3 shows only the regions with large charge density changes ($|\Delta\rho| > 0.06$ e/Å$^3$). The charge transfer in the low-density region (e.g., the charge transfer from above to below the graphene sheet) also contributes significantly to the interfacial electric field. In fact, the calculated induced surface dipole moment, i.e., the change of the surface dipole moment upon graphene adsorption, points towards outside the surface, consistent with the observed reduction of the work function.

Having discussed the electronic structure at the graphene/Ru(0001) interface, we now explain why the hump is formed in the 3.0-nm superstructure. Just matching the (12×12) graphene primitive cells to the (11×11) Ru(0001) primitive cells, the graphene lattice is under ~1.3% tensile strain in the experimental 3.0-nm superstructure and is thus not expected to have a significant buckling as observed by both experiments and our calculations. Our results above have showed that the interfacial C-Ru bonding of the C atoms occupying the atops sites on Ru(0001) plays an important role in dictating the structure and energetics at the graphene/Ru(0001) interface. However, the graphene/Ru lattice mismatch prevents every carbon atom from occupying the atop site. Instead, only part of the carbon atoms reside approximately at the atop positions, giving rise to a 3% C-C bond stretching in these areas. The resulted compressive strain to the rest of the graphene sheet causes significant buckling and the formation of the humps of ~1.7 Å in height. Thus, it is the preference of the direct C-Ru bonding at the atop sites that results in a distribution of tensile and compressive strains in the graphene sheet, causing the corrugated graphene as observed in experiments.

We now turn our attention to the 2.7-nm superstructure. From a pure geometrical perspective, the best match between graphene and Ru(0001) is graphene (11×11) matched to Ru(0001) (10×10) (~0.1% misfit based on experimental lattice parameters or ~0.5% misfit based on computed lattice parameters), leading to the 2.7-nm superstructure. Fig. 5 shows our optimized structure for this superstructure. Compared with the 3.0-nm superstructure, one can see that the 2.7-nm superstructure also has a large buckling within the graphene sheet and the flat region of the graphene sheet is ~2.3 Å above the substrate. Both the spacing and the buckling are 0.1 Å larger than those in the 3.0-nm superstructure. The top view of the 2.7-nm superstructure shows the same feature that the flat region has more C atoms atop Ru while C atoms in the hump are located over the hollow sites.

Comparing interfacial adhesion, we found that the 2.7-nm superstructure has an adhesion energy of 2.7 eV per superstructure (11 meV/C-atom or 0.07 J/m$^2$), about half the strength of the 3.0-nm superstructure. This result indicates that the 3.0-nm superstructure is indeed thermodynamically more stable, supporting the fact that more cases of the 3.0-nm superstructure have been reported and up to millimeter-scale of the 3.0-nm superstructure can be grown.[12,13,15,16] The stronger adhesion for the 3.0-nm superstructure can again be explained in terms of interfacial C-Ru bonding. Just inspecting the top views of the two superstructures [Fig. 1(b) and Fig. 5(b)], one can tell that the 3.0-nm superstructure has more C atoms at the atop sites. If one quantitatively defines interfacial C-Ru bonds as $r_{C-Ru} < 2.3$ Å, there are many more C-Ru bonds in the 3.0-nm superstructure than in the 2.7-nm superstructure (37 versus 10). The larger spacing between graphene and Ru(0001) and the greater buckling within graphene in the 2.7-nm superstructure also indicate weaker interfacial strength and larger strain within graphene, compared to the 3.0-nm superstructure.. Apparently, the 3.0-nm superstructure strikes the right balance between minimizing strain within graphene and maximizing the interfacial bonding, so it is thermodynamically more stable.

The buckling of the graphene sheet we found here due to C-Ru bonding can explain well the aromatic hemispheres formed on Ru(0001),[29] where the edge carbons of the aromatic molecules presumably "grasp" the Ru atoms, causing the center to rise under the compressive strain. Moreover, the nm-sized periodic humps within the graphene layer can serve as a good template for making metal nanoclusters.[10] Further, we have demonstrated that the graphene lattice is well preserved despite the buckling of the graphene sheet and its chemical interaction with the substrate

## IV. SUMMARY AND CONCLUSIONS

We have studied the graphene/Ru(0001) interface with the first-principles DFT method at the GGA-PBE level. By modeling the experimentally observed 3.0-nm and 2.7-nm superstructures, we obtained the optimized interfacial structure and found that a relatively high (~1.7 Å) hump appears in the graphene sheet. The relatively flat region of graphene is found to be 2.2-2.3 Å above the substrate surface, indicating stronger interaction between graphene and Ru(0001) than just physisorption. We found that the interfacial C-Ru bonding arises from the C atoms occupying the atop sites in the flat region. Both electron density difference and local density of states plots show that charge transfer and orbital hybridization concentrate on these interfacial C-Ru bonds. In the hump region, the C atoms are located over the hollow sites. The computed work function for Ru decreases significantly upon graphene adsorption, in agreement with experiment, and also indicates chemical interaction at the interface. We found that the 3.0-nm superstructure has an interfacial adhesion energy about twice of that for the 2.7-nm superstructure, indicating that the 3.0-nm superstructure is thermodynamically more stable. We conclude that the buckling in the graphene layer and the strong interaction (beyond just physisorption) between graphene and Ru(0001) are due to C-Ru bonding at the interface. The 3.0-nm superstructure maximizes such interfacial C-Ru bonding while minimizing lattice mismatch, therefore more stable. This result may help explain and understand the growth of graphene on Ru(0001), which has the potential to produce macroscopic scale of graphene for device applications.

**Acknowledgement**

This work was supported by Office of Basic Energy Sciences, U.S. Department of Energy under Contract No. DE-AC05-00OR22725 with UT-Battelle, LLC, and by Office of Nonproliferation Research and Development (NA22), U.S. Department of Energy. This research used resources of the National Energy Research Scientific Computing Center, which is supported by the Office of Science of the U.S. Department of Energy under Contract No. DE-AC02-05CH11231.

References and footnotes

[1] K. S. Novoselov, A. K. Geim, S. V. Morozov, D. Jiang, Y. Zhang, S. V. Dubonos, I. V. Grigorieva, and A. A. Firsov, Science **306**, 666 (2004).

[2] A. K. Geim and K. S. Novoselov, Nat. Mater. **6**, 183 (2007).

[3] K. S. Novoselov, A. K. Geim, S. V. Morozov, D. Jiang, M. I. Katsnelson, I. V. Grigorieva, S. V. Dubonos, and A. A. Firsov, Nature **438**, 197 (2005).

[4] Y. B. Zhang, Y. W. Tan, H. L. Stormer, and P. Kim, Nature **438**, 201 (2005).

[5] M. I. Katsnelson, K. S. Novoselov, and A. K. Geim, Nat. Phys. **2**, 620 (2006).

[6] K. S. Novoselov, Z. Jiang, Y. Zhang, S. V. Morozov, H. L. Stormer, U. Zeitler, J. C. Maan, G. S. Boebinger, P. Kim, and A. K. Geim, Science **315**, 1379 (2007).

[7] X. S. Wu, X. B. Li, Z. M. Song, C. Berger, and W. A. de Heer, Phys. Rev. Lett. **98**, 136801 (2007).

[8] X. Wang, Y. Ouyang, X. Li, H. Wang, J. Guo, and H. Dai, Phys. Rev. Lett. **100**, 206803 (2008).

[9] C. Berger, Z. M. Song, X. B. Li, X. S. Wu, N. Brown, C. Naud, D. Mayo, T. B. Li, J. Hass, A. N. Marchenkov, E. H. Conrad, P. N. First, and W. A. de Heer, Science **312**, 1191 (2006).

[10] A. T. N'Diaye, S. Bleikamp, P. J. Feibelman, and T. Michely, Phys. Rev. Lett. **97**, 215501 (2006).

[11] J. Coraux, A. T. N'Diaye, C. Busse, and T. Michely, Nano Lett. **8**, 565 (2008).

[12] S. Marchini, S. Günther, and J. Wintterlin, Phys. Rev. B **76**, 075429 (2007).

[13] Y. Pan, D. X. Shi, and H. J. Gao, Chin. Phys. **16**, 3151 (2007).

[14] A. L. Vázquez de Parga, F. Calleja, B. Borca, M. C. G. Passeggi, J. J. Hinarejos, F. Guinea, and R. Miranda, Phys. Rev. Lett. **100**, 056807 (2008).

[15] P. W. Sutter, J. I. Flege, and E. A. Sutter, Nat. Mater. **7**, 406 (2008).

[16] Y. Pan, N. Jiang, J. T. Sun, D. X. Shi, S. X. Du, F. Liu, and H.-J. Gao, Adv. Mater., in press (Doi: 10.1002/adma.200800761; arXiv:0709.2858v1).

[17] B. Wang, M. L. Bocquet, S. Marchini, S. Günther, and J. Wintterlin, Phys. Chem. Chem. Phys. **10**, 3530 (2008).

[18] G. Kresse and J. Furthmüller, Phys. Rev. B **54**, 11169 (1996).

[19] G. Kresse and J. Furthmüller, Comput. Mater. Sci. **6**, 15 (1996).

[20] J. P. Perdew, K. Burke, and M. Ernzerhof, Phys. Rev. Lett. **77**, 3865 (1996).

[21] G. Held, W. Braun, H.-P. Steinrück, S. Yamagishi, S. J. Jenkins, and D. A. King, Phys. Rev. Lett. **87**, 216102 (2001).

[22] D. E. Jiang, B. G. Sumpter, and S. Dai, J. Am. Chem. Soc. **128**, 6030 (2006).

[23] P. E. Blöchl, Phys. Rev. B **50**, 17953 (1994).

[24] G. Kresse and D. Joubert, Phys. Rev. B **59**, 1758 (1999).

[25] G. Bertoni, L. Calmels, A. Altibelli, and V. Serin, Phys. Rev. B **71**, 075402 (2005).

[26] T. C. Leung, C. L. Kao, W. S. Su, Y. J. Feng, and C. T. Chan, Phys. Rev. B **68**, 195408 (2003).

[27] A. Michaelides, P. Hu, M.-H. Lee, A. Alavi, and D. A. King, Phys. Rev. Lett. **90**, 246103 (2003).

[28] L.-L. Wang and H.-P. Cheng, Phys. Rev. B **69**, 045404 (2004).

[29] K. T. Rim, M. Siaj, S. X. Xiao, M. Myers, V. D. Carpentier, L. Liu, C. C. Su, M. L. Steigerwald, M. S. Hybertsen, P. H. McBreen, G. W. Flynn, and C. Nuckolls, Angew. Chem.-Int. Edit. **46**, 7891 (2007).

Captions:

FIG. 1 (color online). The 3.0-nm superstructure of graphene/Ru(0001) with relative height (H) of C atoms within the graphene lattice coded by color: (a) side view of the optimized superstructure; (b) top view of the superstructure (the top Ru layer in light gray, and the subsurface Ru layer in dark gray).

FIG. 2. A perspective view of humps in the graphene layer on Ru(0001) (Ru not shown).

FIG. 3 (color online). Electron density difference between the graphene/Ru(0001) superstructure and the isolated graphene sheet and Ru (0001) surface: (a) side view and (b) top view of the superstructure. Lines: black, graphene lattice; green, Ru lattice. Isosurfaces: blue, accumulation; red, depletion. Isovalue: 0.06 e/Å$^3$. Squares indicate zoomed-in regions.

FIG. 4 (color online). Local density of states for a C-Ru bond where the C atom is atop the Ru atom (see the bond indicated by the straight arrow in Fig. 4b).

FIG. 5 (color online). The 2.7-nm superstructure of graphene/Ru(0001) with relative height (H) of C atoms within the graphene lattice coded by color: (a) side view of the optimized superstructure; (b) top view of the superstructure (the top Ru layer in light gray, and the subsurface Ru layer in dark gray);

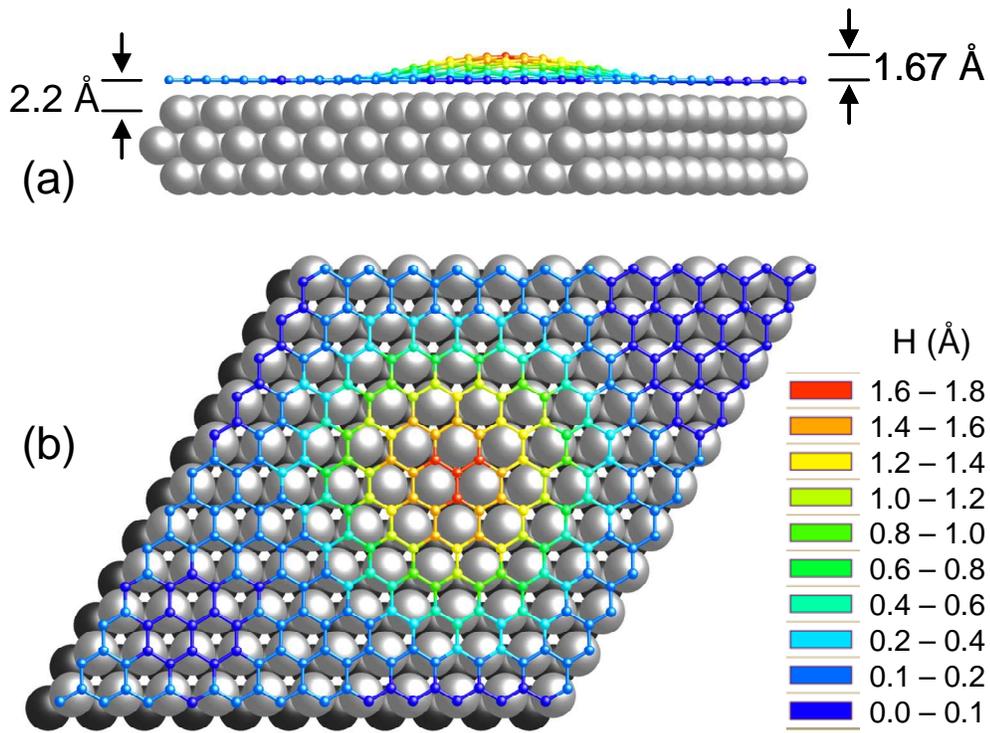

FIG. 1 (color online). The 3.0-nm superstructure of graphene/Ru(0001) with relative height (H) of C atoms within the graphene lattice coded by color: (a) side view of the optimized superstructure; (b) top view of the superstructure (the top Ru layer in light gray, and the subsurface Ru layer in dark gray).

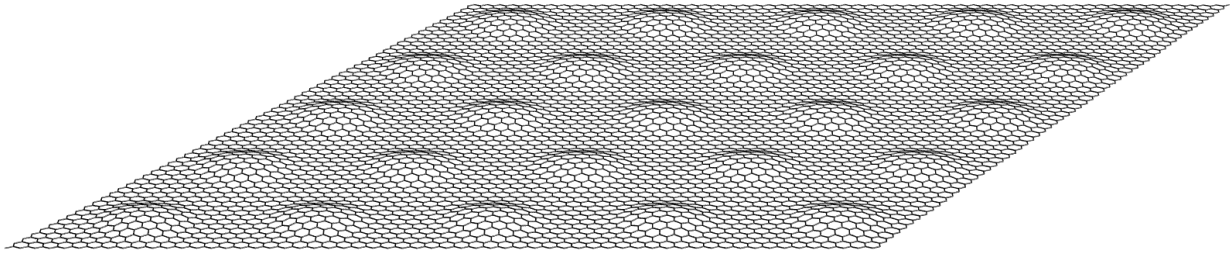

FIG. 2. A perspective view of humps in the graphene layer on Ru(0001) (Ru not shown).

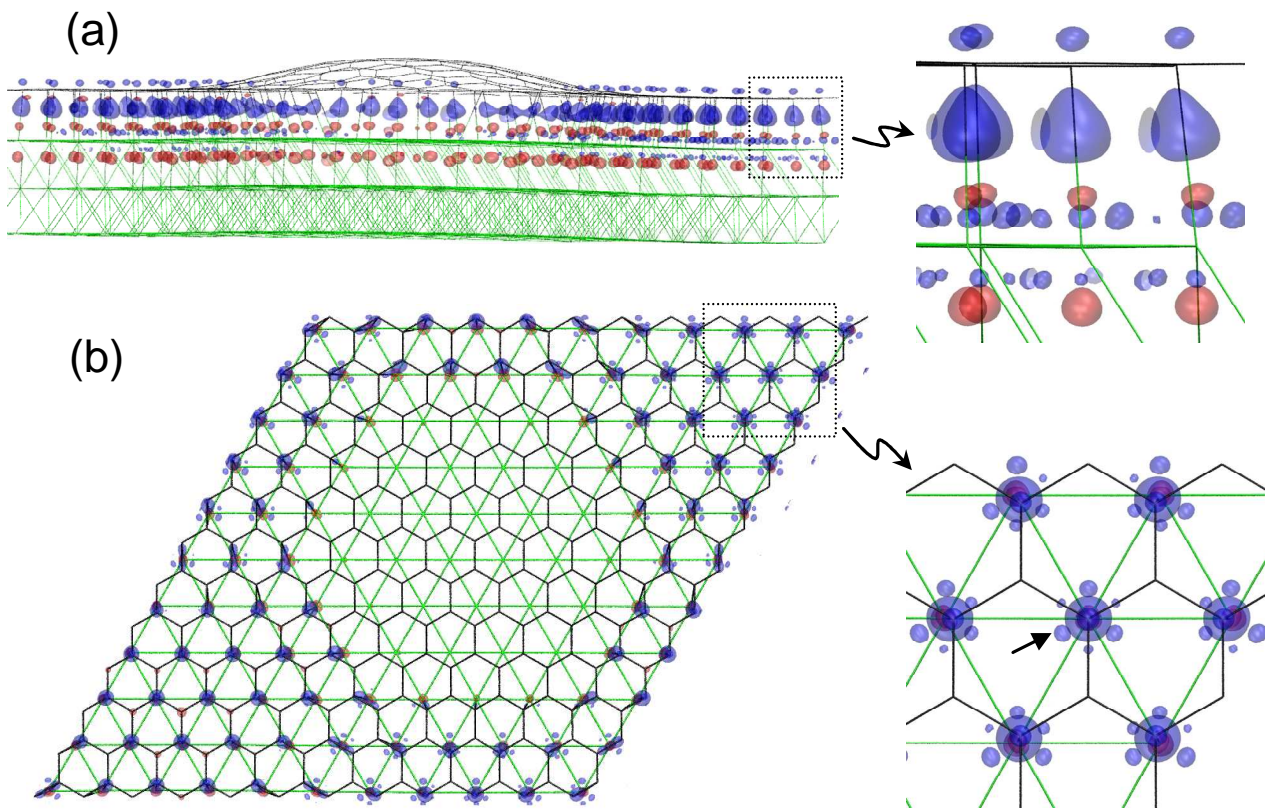

FIG. 3 (color online). Electron density difference between the graphene/Ru(0001) superstructure and the isolated graphene sheet and Ru (0001) surface: (a) side view and (b) top view of the superstructure. Lines: black, graphene lattice; green, Ru lattice. Isosurfaces: blue, accumulation; red, depletion. Isovalue: 0.06 e/Å$^3$. Squares indicate zoomed-in regions.

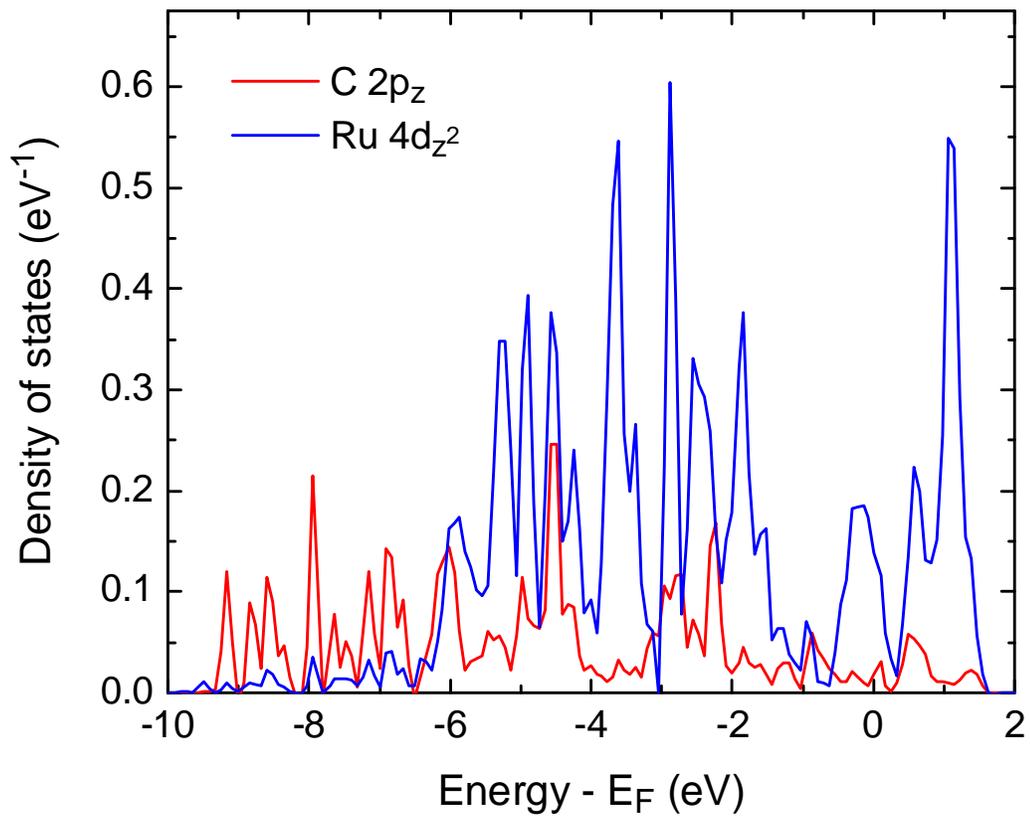

FIG. 4 (color online). Local density of states for a C-Ru bond where the C atom is atop the Ru atom (see the bond indicated by the straight arrow in Fig. 4b).

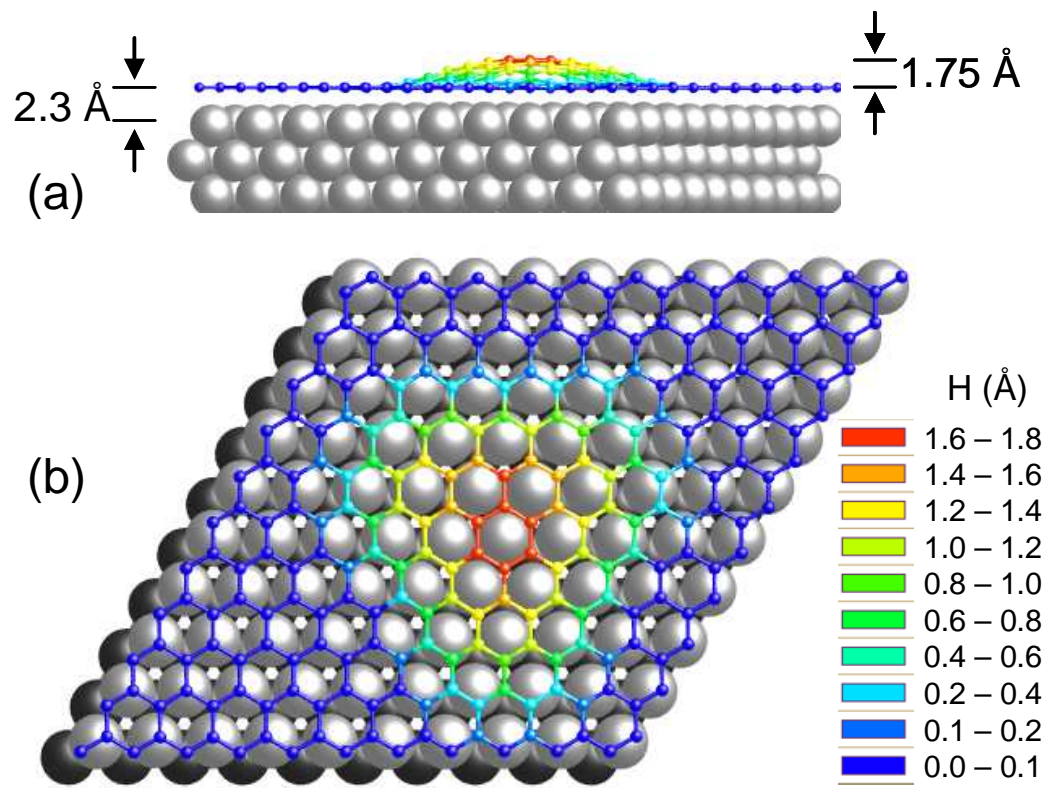

FIG. 5 (color online). The 2.7-nm superstructure of graphene/Ru(0001) with relative height (H) of C atoms within the graphene lattice coded by color: (a) side view of the optimized superstructure; (b) top view of the superstructure (the top Ru layer in light gray, and the subsurface Ru layer in dark gray);